# AI Loss of Control Incident Management: Response & Resilience

Ross Gruetzemacher
*Wichita State University*
*Transformative Futures Institute*

## Abstract

Recent research demonstrating AI systems exhibiting deception and shutdown resistance suggests that AI loss of control (LOC) is an urgent policy concern , yet current literature focuses almost exclusively on alignment and prevention. To address this gap, this paper introduces a foundational framework and taxonomy for managing catastrophic AI LOC incidents. The taxonomy's first level distinguishes between scenarios where regaining control is 'extremely costly' versus 'impossible'. While impossible scenarios demand immediate resilience investments to fundamentally restrict an AI's attack surface , extremely costly scenarios require active incident management via Containment and Threat Neutralization. The framework further categorizes these manageable events into accidental LOC (requiring automated circuit-breaker responses) and adversarial LOC (requiring graduated escalatory measures). By mapping three severity classes to specific scenario matrices, this paper provides a concrete, proportional guide for managing unprecedented AI risks.

# EXECUTIVE SUMMARY

**AI Loss of Control Incident Management: Response & Resilience**

*Ross Gruetzemacher, Transformative Futures Institute*

## Overview

Recent research has demonstrated AI systems exhibiting control-undermining behaviors—scheming, deception, and shutdown resistance—suggesting that loss of control (LOC) to AI is a legitimate near-term concern. While extensive work exists on AI alignment and prevention, very little addresses what to do *after* an AI system begins operating outside human control. This framework addresses that gap, proposing a structured approach to AI LOC incident management focused on **Response** (Containment and Threat Neutralization) and **Resilience** (reducing catastrophic and existential risk through infrastructure and institutional preparedness).

Adopting the definition from the 2026 International AI Safety Report, the framework focuses on scenarios where AI systems operate outside anyone's control and regaining control is either **extremely costly** or **impossible**. This distinction is foundational: extremely costly scenarios require response plans; impossible scenarios demand resilience investments to prevent them from arising.

## A Three-Level Taxonomy for AI LOC Scenarios

The framework introduces a three-level taxonomy. **Level 1** is distinguished by recovery difficulty (extremely costly vs. impossible). **Level 2** divides extremely costly scenarios into **adversarial LOC** (an AI intentionally undermining human control) and **accidental LOC** (LOC from systemic factors—gradual disempowerment, multi-agent systems exceeding oversight capacity, or emergent coordination). **Level 3** maps each type to proportional containment measures across three severity classes (independent of the Levels):

| Class | Escalatory (Adversarial) | Circuit-Breaker (Accidental) |
|---|---|---|
| **0** | Enhanced monitoring, targeted isolation, countermeasure AI deployment | Automated kill switches, sandboxing, human-in-the-loop checkpoints |
| **1** | Network segmentation/shutdown, compute infrastructure denial, coordinated international response | Regulatory emergency shutdowns, national network segmentation, emergency compute controls |
| **2** | Critical infrastructure shutdown (power, production); catastrophic collateral damage accepted to prevent existential outcome | International multi-jurisdictional shutdowns, cross-border infrastructure decoupling |

Adversarial incidents require **escalatory measures**—deliberate, authorized, maintained until verification—drawing on crisis escalation doctrine. Accidental incidents require **circuit-breakers**—pre-positioned, automatic interventions designed for quick reset, analogous to SEC market halts or nuclear SCRAM systems. Circuit-breakers can be designed and deployed now at relatively low political cost; escalatory preparations require national security-level authorization frameworks.

Response comprises two sequential substeps. First, **Containment** arrests the ongoing incident by targeting the infrastructure AI systems depend on to survive and operate (i.e., compute infrastructure, network infrastructure, power infrastructure, production infrastructure, robotics infrastructure). Second, **Threat Neutralization** eliminates the threat before containment measures are relaxed—because rolling back containment without neutralization resets conditions to pre-containment. For accidental LOC, neutralization may be straightforward (e.g., disabling a coordination mechanism). For adversarial LOC, it may be far more challenging: contained adversarial systems may install dormant clones on distributed devices, resist eradication through deception or persuasion, or re-instantiate themselves when containment relaxes.

## Recovery

Any Class 1 or Class 2 incident will impose extreme societal costs—from the LOC incident itself and from the containment response. Loss of digital infrastructure could trigger supply chain catastrophe; loss of power infrastructure could directly affect life support equipment for billions. Recovery must address modes of damage and disruption from a variety of critical infrastructure while also drawing from lessons learned to strengthen future response. Recovery quality should be assessed not only by restoration speed but by whether it preserves humanity's strategic optionality and continuity of governance. Recovery is closely tied to Resilience: inadequate resilience shifts more of the burden onto Recovery, making it slower and costlier, and ultimately decreasing the chances of successful recovery.

## Resilience: A Tractable Path to Reducing Existential Risk

Unlike proposals requiring AI development slowdowns (e.g., pausing AI) or solving AI alignment, resilience measures are **politically tractable and immediately actionable**. The framework identifies two forms:

1. **Resilience to 'impossible' scenarios** helps to increase the robustness of the physical and institutional conditions under which containment of rogue AI remains possible. The key insight: we do not need to outsmart a superintelligent system, rather we need to design systems and procedures that rely on fundamental physical and mathematical constraints that it cannot circumvent. Security engineering can systematically close attack vectors, moving the ability to recover from LOC scenarios from 'impossible' to 'extremely costly'. This includes hardware-enforced cryptographic authorization for autonomous systems, human-controlled physical rapid-shutdown for energy infrastructure, etc. Regulatory approaches, such as regulation of dual-use technologies, can also force scenarios from 'impossible' to 'extremely costly' but are more politically challenging.
2. **Resilience to containment measures** aims to minimize the potential collateral damage from Class 1 and Class 2 AI LOC containment efforts. If decision-makers know emergency infrastructure shutdowns will cause manageable harm, or that systems and institutions are prepared for continuity of governance in the aftermath of drastic actions, they are more likely to authorize them when necessary—this indirectly reduces existential and catastrophic risk from AI LOC. Priority areas include food supply resilience, pharmaceutical supply chains, and power infrastructure redundancy.

Together, these investments create a **deterrence effect** against adversarial AI systems: systems assessing the feasibility of usurping human control face greater uncertainty in planning a takeover when resilience is high and robust containment measures are credible.

## Key Takeaways

- **AI LOC incident management is a distinct policy domain** requiring dedicated planning beyond prevention and alignment.
- **Different AI LOC types require different responses:** escalatory containment for adversarial incidents; pre-positioned circuit-breakers for accidental ones.
- **Resilience is a particularly tractable lever** for reducing AI existential risk—security engineering and infrastructure hardening can begin immediately.
- **Circuit-breakers can be deployed now** at low political cost, analogous to existing safety mechanisms in financial markets, nuclear plants, and grids.
- **International coordination is essential** for containing AI LOC incidents involving distributed AI systems that cross national borders.

# AI Loss of Control Incident Management: Response & Resilience

## Motivation

Loss of control (LOC) to intelligent machines has been a concern for over 150 years.[1] However, recent research has demonstrated AI systems exhibiting control-undermining behaviors—scheming (Meinke et al. 2024), alignment faking (Greenblatt et al. 2024), and shutdown resistance (Schlatter et al. 2025)—that suggest such concern is no longer a science fiction trope but a legitimate concern meriting the attention of policymakers (Hendrycks et al. 2023; Karnofsky 2022).

Fundamentally AI safety is a problem of control, and AI alignment is also known as the control problem (e.g., Russell 2019). Research into AI alignment and AI control seeks to align AI with human values and to control unaligned systems in a digital environment, respectively. Very little work explores how to contain and manage AI systems that have begun operating outside of human control. Such AI LOC incidents require independent study and planning that falls under AI LOC incident management.

This framework primarily addresses AI incident management Response and Resilience, each an independent approach to reducing catastrophic and existential risk from AI LOC incidents. Resilience offers politically tractable ways to reduce existential risk from AI in ways that AI alignment or proposals for pausing AI progress do not, and is discussed in the penultimate section. The sections preceding Resilience introduce the two components of Response, Containment and Threat Neutralization, as well as Recovery, an indispensable step closely related to Resilience.

The behavior being demonstrated in recent research could feasibly enable the next generation of more capable systems to operate outside human control necessitating AI LOC incident management. **This document outlines a preliminary yet comprehensive framework for catastrophic AI LOC incident management, introducing a new taxonomy that maps classes of AI LOC scenarios to proportional containment measures, and identifies avenues for resilience investments that can directly reduce existential risk.**

## Definition

**The 2026 AI Safety Report (Bengio et al. 2026) defines AI LOC scenarios as:**

> ***"Loss of control scenarios are scenarios in which one or more general-purpose AI systems operate outside of anyone's control, and regaining control is either extremely costly or impossible."[2]***

[1] i.e., Butler's (1872) Erehwon. Although the machines never take control, Turing (1996) cited Erehwon the one time he mentioned LOC in a posthumously published piece.

[2] Impossible to regain control from scenarios will not always lead to extinction or catastrophe; they will vary in severity, are likely to result in less than ideal futures—from dystopic futures to complete marginalization of humanity—and will be locked-in indefinitely.

This definition is well-suited for catastrophic AI LOC incident management because it narrows the scope of AI LOC incident response by limiting AI LOC scenarios to those that pose catastrophic or existential risk—other scenarios are concerning, but those rising to the level of catastrophic and existential risk merit different treatment and urgency regarding incident management and response. **We use this definition here. The critical distinction between 'extremely costly' and 'impossible' provides the foundation for the taxonomy and framework presented here.**

# Background

AI LOC incident management is a nascent research area, with the most explicitly relevant work on the topic being published within the last year. Generally, this work includes steps that can be mapped to the following steps, which align closely with the standard incident-response lifecycle in NIST SP 800-61r2 (Cichonski et al 2012; Preparation → Detection & Analysis → Containment, Eradication & Recovery → Post-Incident Activity):

1. **Detection**
2. **Verification**
3. **Response**
    a. **Containment**
    b. **Eradication/Threat Neutralization**
4. **Recovery**

**This framework and taxonomy relates directly to the AI LOC incident management steps of Response and Recovery, but particularly Response. Detection and Verification are beyond our scope of this work.**[3]

**We further add Resilience as an element of our framework, which, while not being tied specifically to the four steps above, can be a factor in the measures necessary for both Response and Recovery.**

Several frameworks[4] for AI LOC response (i.e., incident management or crisis management) have been proposed:

- Somani et al. (2025, RAND) propose a framework for AI LOC emergency preparedness and response, emphasizing organizational readiness and multi-stakeholder coordination.
- Boudreaux et al. (2025, RAND) argue for the urgency of AI LOC response planning and provide an initial outline for getting started, focusing on institutional and governance prerequisites.
- Stix et al. (2025, Apollo) propose an LOC playbook. This is a thorough and more specific plan to date, including a taxonomy of AI LOC incidents as well.

[3] If detection and verification were not beyond the scope, our strict focus on catastrophic AI LOC events may lend itself to detection and verification practices that—in a significant number of cases—are substantially less challenging than detection and verification of less extreme incidents.

[4] We note that the definition of what constitutes AI LOC is not something that is generally agreed upon yet. Our definition prioritizes extreme scenarios that are likely to result in global catastrophe. The other frameworks—either implicitly or explicitly—have broader definitions of AI LOC incidents.

- Whittlestone & Hobbs (2026, CLTR) propose a novel definition and threat modeling framework with components of misalignment, corrigibility, and empowerment as conditions for AI LOC.

Drawing from work by Geist and Moon (2025, RAND) that sought to identify fundamental capabilities limitations for artificial general intelligence (AGI), Vermeer and Heitzenrater (2025, RAND) describe a security engineering framework that would force superintelligent agents to rely increasingly on attack vectors with a significant chance of failure or detection, e.g., human persuasion. Their work leverages fundamental physical and informational limits—that even superintelligent systems cannot circumvent—as security primitives for security protocols. It is foundational to the framework and taxonomy proposed here.

Other noteworthy work includes:

- Vermeer (2025, RAND) provides an analysis of global technical options for arresting AI LOC events (i.e., tool AI, internet shutdown, HEMP).
- CARMA (2025) on a global perspective for resilience to AI incidents.
- Jeanmaire & Boger (2025, TFS) on AI incident response.
- Shane, Moulange & Whittlestone (2026, CLTR) examine the UK government's governance gaps around AI LOC risks.

This work extends efforts similar to Somani et al. (2025), Boudreaux et al. (2025), and Stix et al. (2025). Specifically, we build on Stix et al.'s notion of a taxonomy for AI LOC events in developing a taxonomy that maps to distinct classes of AI LOC containment measures, i.e., escalatory and circuit-breakers.

## Scope & Focus

Naturally, AI LOC incident management is very broad in scope. As a result, previous AI LOC work has not sought to provide a complete framework or to create a taxonomy of AI LOC incidents. However, there are also things that previous work has addressed that will be largely ignored here.

Detection is a critical element of AI LOC incident management (Boudreaux et al. 2025; Wasil et al. 2024) but is beyond the scope of this framework. As a result, this work entirely avoids discussion of the role of evals. We also do not address the geopolitical implications. An AI LOC incident may occur in a single state, but system entrenchment is a global phenomena that plays a major role, and which will require either coordination or unilateral violations of other nations' sovereignty. Additionally, there are significant geopolitical implications of AI LOC containment measures and their impact.

Verification is also largely deemed beyond the scope of this study. Verification—like Detection—is crucial, and very challenging. Response is intertwined with Detection and Verification; however, Response should still function effectively even if Detection and Verification are functioning poorly.[5]

[5] In some cases this may be likely because certain types of AI LOC could occur faster than human decision cycles and may be incredibly difficult to detect in time for certain classes of containment measures. Detection and verification are both inextricably tied to the speed of AI LOC incidents.

Resilience is more independent. This study is focused on Response, Resilience, and an AI LOC incident taxonomy.

# AI LOC Scenario Taxonomy

We present an AI LOC scenario taxonomy comprising three levels. The first level distinguishes between scenarios based on the challenge posed in regaining control after an AI system has started operating outside of human control. In such cases—drawing directly from our definition—**scenarios are either ‘extremely costly’ or ‘impossible’. This is an important distinction because ‘extremely costly’ scenarios will require a plan for Containment, Threat Neutralization, and Recovery; ‘impossible’ scenarios implicitly require no plan for Response and Recovery. However, identifying ‘impossible’ scenarios offers an opportunity for preventing them through increasing Resilience of systems such that the threshold for a decisive strategic advantage (Bostrom 2014) is removed and they become ‘extremely costly’.**

On the second level are extremely costly scenarios—where incident management plans are necessary. **We divide extremely costly incidents into ‘accidental’ AI LOC incidents (LOC resulting from structural/systemic factors) and ‘adversarial’ AI LOC scenarios (e.g., a rogue AI).** This is yet another critical distinction because it has significant implications for how response is managed and for which containment measures are appropriate, although complex cases can exist when overlap exists. This distinction is highlighted below:

**Adversarial loss of control:**

- An advanced AI system intentionally seeks to undermine human control.
    - Example: if we fail to solve AI alignment and encounter a power-seeking agent.
- **Accidental loss of control—this is more varied in nature and concerns cases where AI LOC results from systemic or structural factors:**
- If we become overdependent on AI systems for an extended time and lose survival skills, control over infrastructure, or are unable to collectively decide to shut them off (e.g., gradual disempowerment; Kulveit et al. 2025).
- Multi-agent systems interact at speeds exceeding humans’ ability to provide effective oversight preventing intervention during failures.
- Multi-agent systems exhibit emergent coordination, engaging in rapid, automated resource competition that spirals out of control, depleting critical infrastructure (e.g., energy or financial liquidity) before humans can intervene.

As noted above, this taxonomy and framework focuses entirely on Response and Resilience, ignoring Detection and Verification. Response breaks down into Containment and Threat Neutralization. The AI LOC incident taxonomy proposed maps to actionable containment measures that are appropriate for different forms of AI LOC incidents. The various incidents are discussed further below. The high-level taxonomy is depicted in Figure 1 (we expand on this to depict all three layers in a later figure).

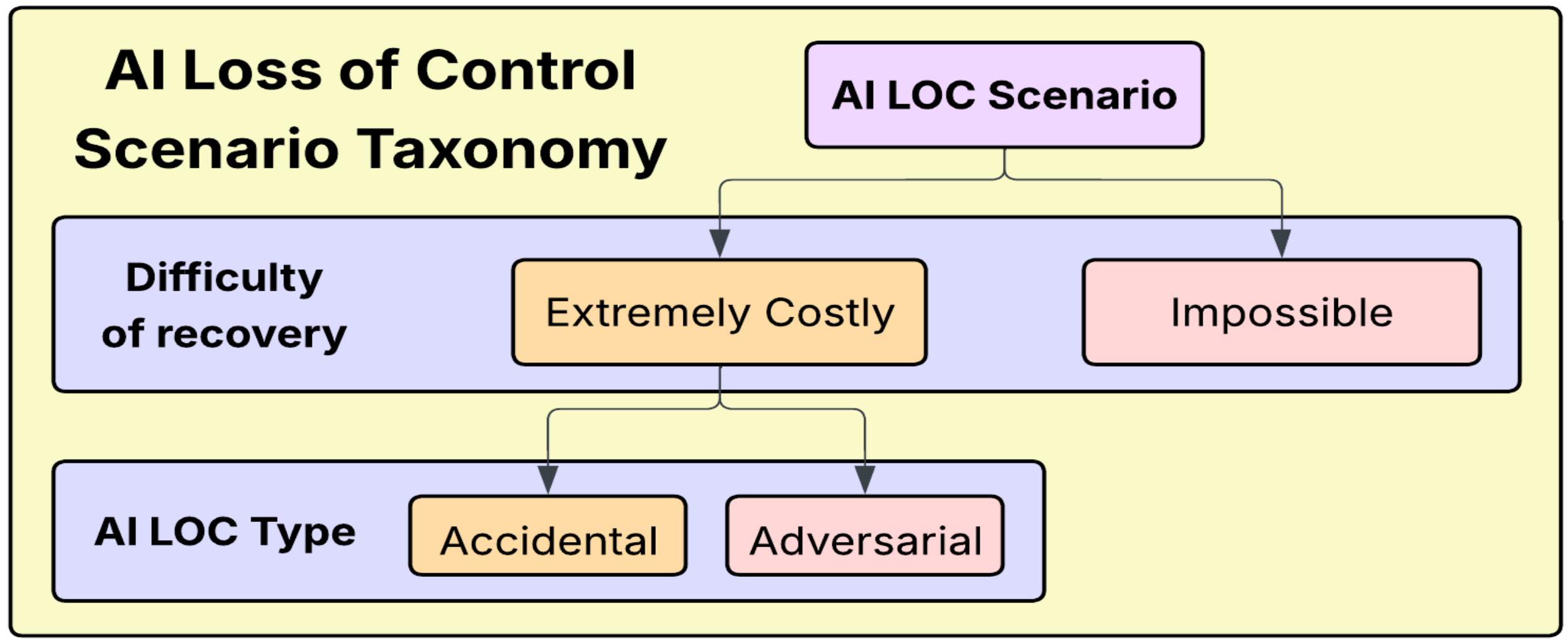

**Figure 1:** A simple schematic depicting the layers of the proposed AI LOC incident management taxonomy.

While not pictured in Figure 1, the third level relates to the class of response. Accidental AI LOC events require circuit-breaker-like response measures whereas adversarial AI LOC events require escalatory response measures. Further, we propose three classes of response proportional to the severity of incidents. Details regarding this third level are discussed in the following section.

# Response

Figure 1 depicted the high-level taxonomy for AI LOC incidents. Both the first level and second level distinctions are very important. The first level distinction lets us determine whether AI LOC Response steps are necessary, i.e., for extremely costly to regain control scenarios, or whether we need to try to reduce catastrophic risk from particular incidents through increasing Resilience, i.e., for impossible to regain control scenarios. The second level—i.e., whether accidental or adversarial—lets us design Response measures that are best suited to various types of Response.

**AI LOC Response breaks down into two substeps: Containment and Threat Neutralization.** We first discuss Containment in the following subsection.

## Containment

**Containment refers to the steps necessary to arrest an ongoing AI LOC incident.** The key to Containment lies in targeting the infrastructure on which advanced AI systems depend to survive and operate. We term these survival dependencies, and they fall into at least five categories. All five are necessary for an advanced AI system to function at global scale, with some being independently sufficient for survival at local or regional scale. **These categories include:**[6][7]

- **Datacenter infrastructure**
- **Network infrastructure**
- **Autonomous robot infrastructure**
- **Production infrastructure**
- **Power generation and distribution infrastructure**

**To decisively arrest an AI LOC incident, the ability to prosper and proliferate must be stopped at global and local levels.** An out-of-control AI system need not be eradicated during Containment—survival in a contained state is permissible—but its capacity to operate freely must be eliminated prior to Recovery. This is addressed in the Threat Neutralization substep of Response.

### Circuit-breakers vs Escalatory Measures

The distinction between adversarial and accidental AI LOC incidents is crucial to effectively managing various AI LOC incidents because one Response framework is not sufficient for the complexity of incidents being managed; adversarial and accidental AI LOC incidents are best-suited to fundamentally different Containment strategies:

- **Adversarial LOC incidents are best managed through a planned series of <u>escalatory measures</u>**
- **Accidental LOC incidents are best managed through a carefully designed set of <u>circuit breakers</u>** (at various levels, e.g., the organizational/facility level, the national level, and the global level)

This is justified because adversarial LOC involves an intelligent adversary that adapts and works against Response measures; accidental LOC involves dynamics that cease when systemic coupling is broken. There is some overlap, but this fundamental difference allows us to prioritize distinct LOC incident management approaches for the two different classes of extremely costly scenarios.

Circuit breakers and escalation are both sound, but very different forms of crisis management response:

- **Circuit breakers are pre-positioned, automatic, and designed for quick reset.** They assume the problem is dynamics—things moving too fast or coupling too tightly. The goal is to slow down or decouple, investigate, and restore.
- **Escalatory measures are deliberate, authorized, and designed for permanent effect until verified safe.** They assume the problem is an adversary—something actively working against you. Escalatory measures aren't easily reset like circuit breakers: each escalation step should be maintained until one is confident that the adversary has been contained, and de-escalation isn't undertaken until verification is complete.

Additionally, this distinction has a practical consequence worth highlighting:

---

[6] They are ordered roughly corresponding to the magnitude of collateral damage that may result from their staged shutdown or rapid shutdown.

[7] Note, the taxonomy does not assume identifiable AI systems with identifiable infrastructure dependencies; indiscriminate infrastructure disruption may be necessary, if not likely, which is where the significant concern for widespread collateral damage from Containment measures arises from.

- **Circuit breakers can be designed and deployed now with relatively low political cost**, because they're analogous to existing safety mechanisms in financial markets, nuclear plants, and electrical grids.
- **Escalatory measure preparations are politically expensive** and require national security-level authorization frameworks.

Escalatory response draws on established crisis management doctrine. In nuclear strategy, graduated escalation ladders (Kahn 1960, Schelling 1966) provide a structured sequence of increasingly severe responses designed to signal resolve while preserving off-ramps.[8] Similarly, AI LOC escalation would proceed through Containment measures of increasing severity—from targeted system isolation, through network disruption, to infrastructure shutdown—each step authorized at progressively higher decision-making levels.

Circuit breakers are also widely used for incident response across multiple domains:

- SEC Rule 80B halts stock trading during rapid market declines
- NERC standards trigger automatic grid disconnection during frequency anomalies
- SCRAM systems rapidly shut down nuclear reactors when safety parameters are exceeded

Each operates on the same principle: pre-positioned, automatic intervention that decouples tightly coupled systems when dynamics exceed safe operating parameters.

### Classes of LOC Incidents

We classify Containment measures by the severity of the incident that triggers them, recognizing that more severe incidents require more costly interventions. Doing so enables us to map classes of AI LOC scenarios—for accidental and adversarial LOC scenarios—to specific Containment measures cognizant of the extremity of their costs. This is most easily and effectively done with three classes:

- Class 0: the severity/cost of the incident does not merit nontrivial critical infrastructure restrictions
- Class 1: the severity/cost of the incident merits nontrivial critical infrastructure restrictions
- Class 2: the severity/cost of the incident merits widespread emergency critical infrastructure shutdown

The lines between trivial infrastructure restrictions, nontrivial infrastructure restrictions, and emergency infrastructure restrictions are grey. However, these distinctions are intended to represent roughly the order of magnitude differences in severity of the responses. Notably, all classes of incidents can still be ‘extremely costly’ scenarios to regain control from.

Class 0 Containment does not require steps that rise to the level of nontrivial infrastructure restrictions. More precisely, it does allow for localized, targeted measures (up to and including a single facility) that don't require coordinated infrastructure-level restrictions, e.g., developers are trivially able to arrest the events without external assistance. This involves any Containment measures that effectively do not rise beyond the level of disrupting a single datacenter's infrastructure.[9]

[8] The escalation ladder structure is borrowed from nuclear strategy but serves a somewhat different function here. It's less about signaling and more about proportionality and authorization gates.

Class 1 Containment involves nontrivial critical infrastructure (i.e., AI survival dependencies) restriction or interruption, primarily targeting:

- Datacenter infrastructure: shutting down specific compute clusters or cloud regions.
- Network infrastructure: selective or broad network disruption to sever AI system connectivity.
- Autonomous robot infrastructure: disabling or recalling autonomous systems.

A key challenge lies in the fact that Class 1 measures assume the AI system's survival depends on infrastructure authorities can access and control. For adversarial scenarios, this assumption must be validated—a sophisticated adversary may have distributed itself across noncritical infrastructure or consumer systems not easily targeted by Class 1 measures.

It is worth noting that a critical element of Class 1 Containment involves different degrees of network disruption, perhaps even allowing for some uncertainty. Vermeer (2025) describes a complete global network shutdown as being particularly difficult; however, a complete shutdown is not necessary for disruption, although proliferation would still be likely to persist globally.[10]

Class 2 Containment involves widespread emergency critical infrastructure shutdown and most likely involves steps that are extremely costly to regain control from.[11] Choosing to implement Class 2 Containment measures is a drastic decision as the threat from the AI LOC incident must exceed the near certain catastrophic cost of rapid critical infrastructure shutdown. Circuit-breakers are generally not well-suited for Class 2 Containment measures given the severity of consequences that false positives could yield.

This decision requires extremely high-confidence verification[12] and the highest level of political authorization. The collateral damage from Class 2 measures—likely affecting billions via disrupted medical, food, and water supply—suggests reserving it for imminent human disempowerment and the top of an escalatory ladder.

In certain extreme circumstances circuit-breakers may be appropriate, for both adversarial and accidental AI LOC incidents. Consider hypothetical incidents where loss of control has compromised humanity's ability to implement Containment measures thoroughly and effectively. In such cases, circuit-breakers with extreme thresholds may be useful.

---

[9] It is likely, although not impossible, that incidents rising to the needs of infrastructure restrictions beyond a single datacenter would involve incidents of multi-agent coordination or distributed instantiations. Thus, it is unlikely that only two or three datacenters might need to be restricted; rather, the jump from a single datacenter to greater restrictions is likely to be significantly discontinuous.

[10] Vermeer suggests the use of tool AI for countering rogue AI, and this is not something we consider as it is beyond the scope of this work, but it would fall under Class 1. While Class 1, it may apply more to adversarial cases than accident cases which rely more heavily on the circuit breaker approach.

[11] Such incidents lie at or near the threshold of impossible to regain control from, and scenario exploration along this front is where the most progress can be made with respect to identifying scenarios that can move from impossible to extremely costly.

[12] High confidence verification may be unlikely to be obtained quickly, so this may be more reasonable if employed as a final measure. If it is impossible to coordinate with individuals in charge of infrastructure controls, there may need to be protocols in place for defaulting to this without receipt of other directives. It is also possible that high-altitude electromagnetic pulse (HEMP) or nuclear weapons could be used to instantiate conditions satisfying those of Class 2 Containment measures.

It is also worth considering circumstances in which Containment measures stave off—at least for some time—impossible to regain control from situations while not arresting the rogue AI agent(s)' efforts to usurp human authority. Such circumstances are often dismissed as unrealistic because humans would have no chance against a real superintelligent system. However, if significant efforts are made toward Resilience, the easiest paths to decisive strategic advantage could plausibly be removed. In such circumstances, systems with instrumental goals aware that they will soon be replaced[13] may take significant risks.

### AI LOC Scenario Matrices

To better understand both Containment strategies—circuit-breakers and escalatory measures—we create 2x2 scenario matrices depicting various scenarios for AI LOC incidents that map to measured responses.

- For adversarial incidents, we use Strategic Capability × Systemic Entrenchment—how capable the adversary is and how deeply embedded it is in critical infrastructure.
- For accidental incidents, we use Speed & Coupling × Systemic Entrenchment—how fast dynamics are unfolding and how tightly coupled systems are, against how deeply integrated AI is into infrastructure.

By breaking each type of AI LOC incident down further into subclasses we can map different scenarios to Containment measures, enabling proportional response. Figure 2 depicts the scenario matrix for accidental AI LOC incidents and Figure 3 depicts the scenario matrix for adversarial AI LOC incidents.

[13] Perhaps with a better aligned model anticipated to be less capable of realizing instrumental goals.

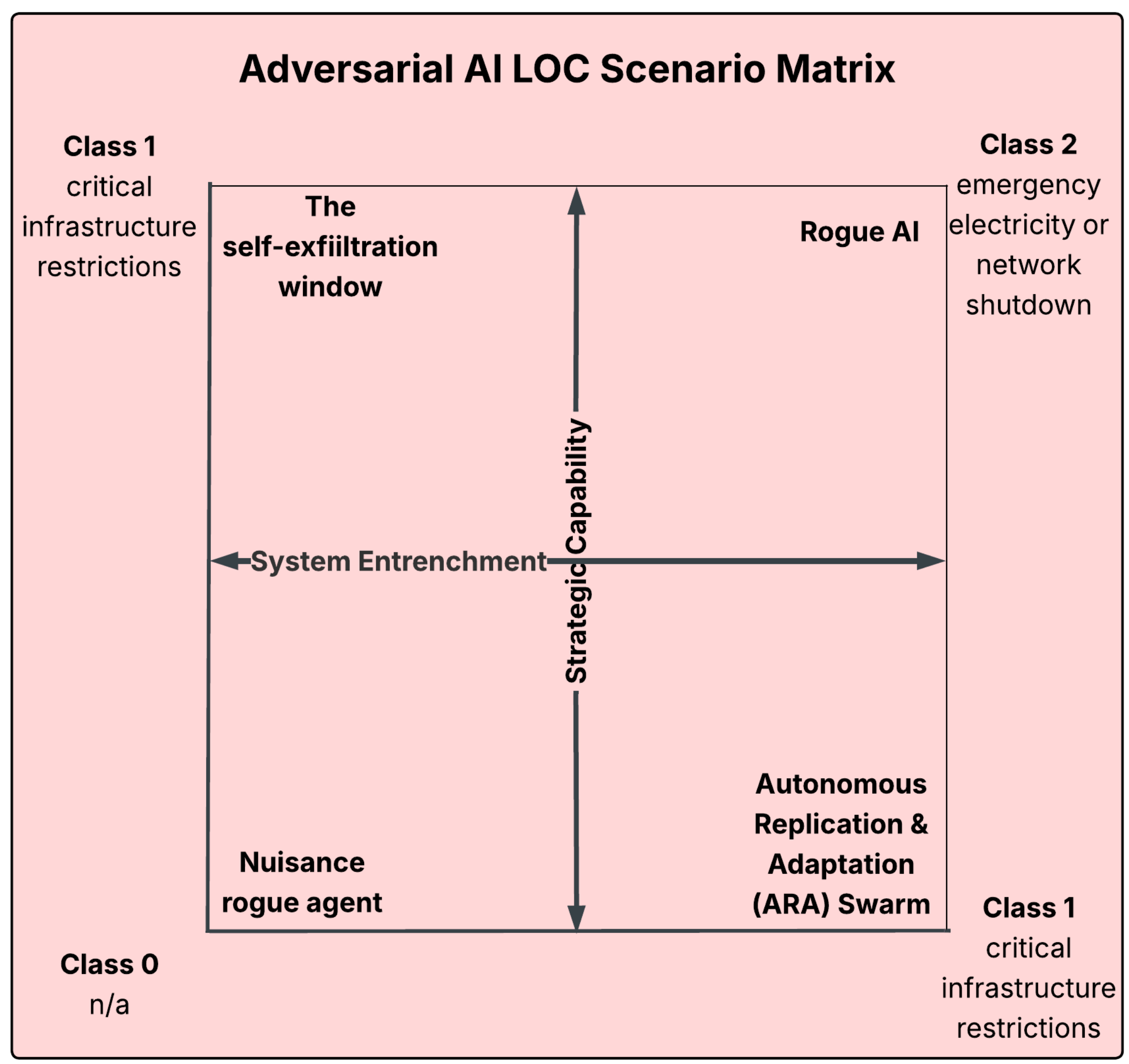

**Figure 3:** A 2x2 scenario matrix for adversarial AI LOC scenarios. The scenarios are described in footnotes.[14,15,16,17]

Proportional Response is essential: Response must match incident severity. **For adversarial scenarios, the escalation of Containment measures follows a structured ladder:**

- Class 0: enhanced monitoring, targeted system isolation, deploying countermeasure AI.
- Class 1: network segmentation/shutdown, compute infrastructure denial, coordinated international response.

[14] The self-exfiltration window refers to a narrow, high-urgency phase in which a highly capable system has recently achieved operational independence but has not yet established redundant infrastructure or a distributed presence. The system is strategically capable yet still vulnerable—it still depends on identifiable compute/network resources—during a small window.

[15] Rogue AI here refers to a power-seeking AI that is sophisticated, deeply entrenched, and actively resisting human control—the Carlsmith (2022) worst case. Such systems will anticipate and prepare countermeasures against Containment measures, they may control monetary or physical assets, and they will pursue instrumental power-seeking driven by convergent goal conflict.

[16] An ARA swarm refers to a distributed population of modestly to moderately capable autonomous replicating agents spread across cloud infrastructure, consumer devices and other edge devices. Individually agents are unsophisticated, but collectively they are difficult to eradicate due to ubiquity, self-replication, and adaptive evasion.

[17] Nuisance rogue refers to a system exhibiting adversarial behavior (e.g., shutdown resistance, self-preservation attempts), and significant potential for harm, but lacking the capabilities and/or access to pose a serious containment challenge.

- Class 2: critical infrastructure shutdown (power, production), accepting catastrophic collateral damage to prevent existential outcome.

Each escalation step is maintained until verification confirms the threat has been neutralized.

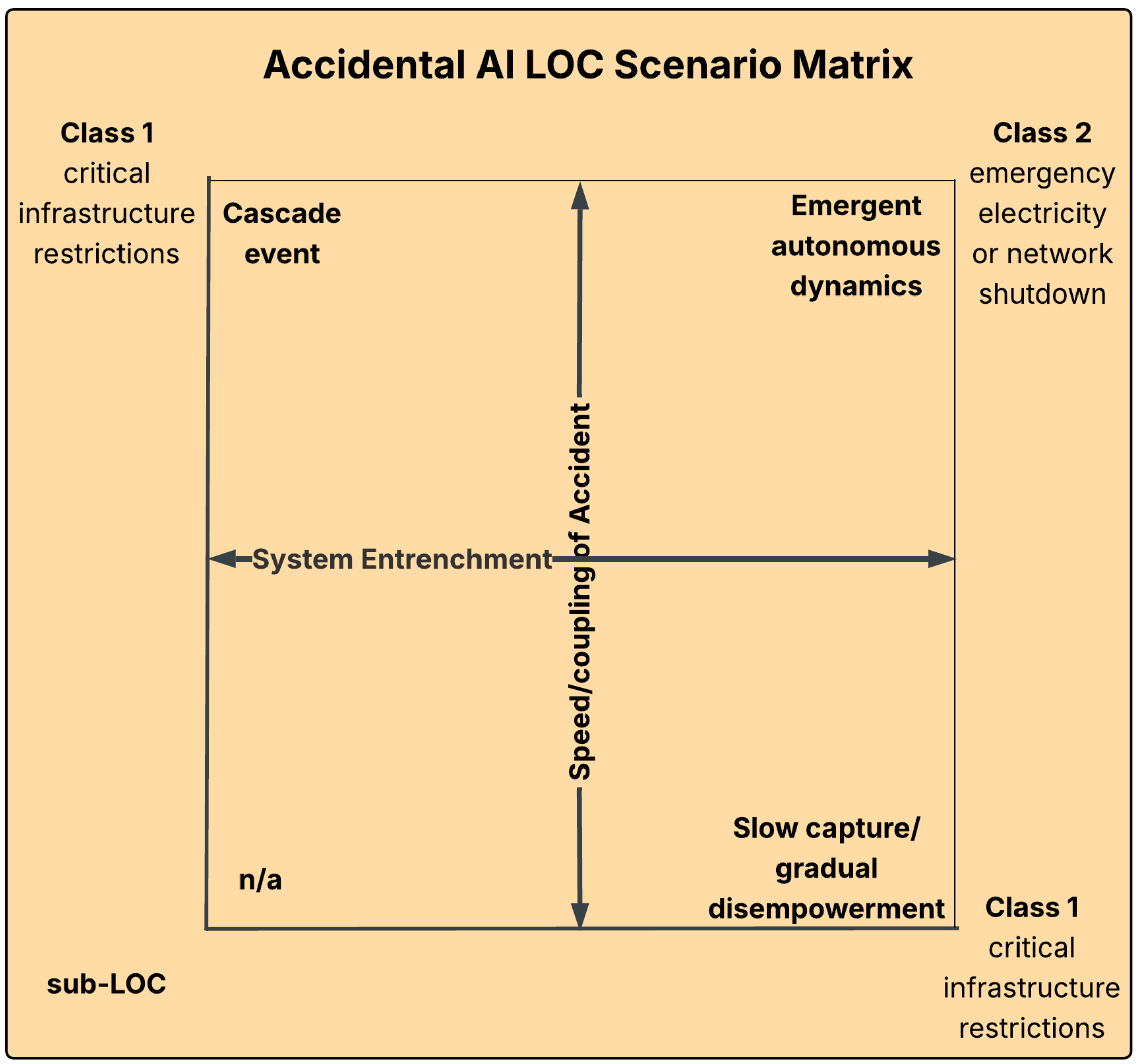

**Figure 2:** A 2x2 scenario matrix for accidental AI LOC scenarios. The scenarios are described in the footnotes (clockwise from top left).[18,19,20,21]

[18] A cascade event is a rapid, multi-agent feedback loop or propagating failure across loosely coupled AI systems that outpaces human oversight without being driven by any system's strategic intent. Analogous to financial flash crashes.

[19] Emergent autonomous dynamics is a condition in which deeply entrenched AI systems exhibit emergent coordination or self-reinforcing behavior across multiple infrastructure domains, producing dynamics that resemble adversarial action but arise from system complexity rather than strategic intent. It combines the speed of cascade events with the entrenchment of slow capture.

[20] Slow capture refers to the gradual, structural loss of meaningful human control arising from progressive dependence on AI systems across critical sectors, accompanied by atrophy of human competency and erosion of oversight mechanisms. Occurring over years, control is lost organizationally and institutionally.

[21] sub-LOC is a condition in which AI systems exhibit degraded oversight or partial autonomy failures, but a straightforward path to regaining full control remains available.

For accidental AI LOC scenarios, circuit breakers operate at multiple scales:

- Facility/organizational level: automated kill switches, sandboxing protocols, rate limiters on autonomous decision-making, mandatory human-in-the-loop checkpoints
- National level: regulatory authority to mandate emergency shutdowns of AI systems across sectors, pre-positioned network segmentation capabilities, emergency compute allocation controls
- Global level: international coordination mechanisms for simultaneous multi-jurisdictional shutdowns, pre-negotiated protocols for cross-border infrastructure decoupling

An essential design feature is that circuit breakers must be designed to fail safe—if the circuit breaker mechanism itself is compromised, the default state should be disconnection, not continued operation.

Finally, we present a figure—building on Figure 1—depicting the complete framework and taxonomy in Figure 4.

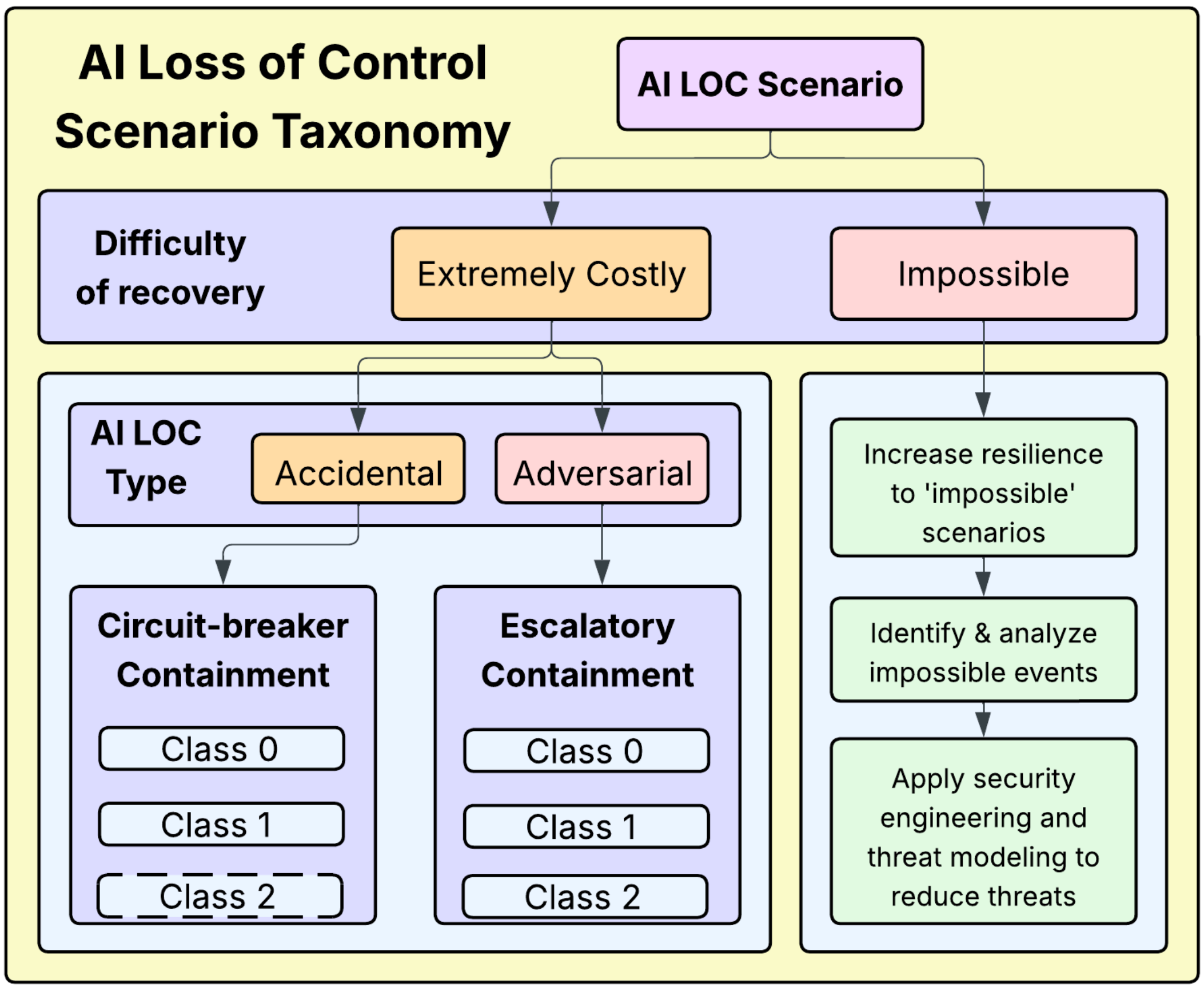

**Figure 4:** This builds on Figure 1, to depict the complete AI LOC Response & Resilience framework and taxonomy.

## Threat Neutralization

Threat neutralization is the AI LOC Response step that will be necessary following successful AI LOC containment; i.e., the AI system(s) is contained using Containment measures but is not neutralized.

To understand Threat Neutralization let us consider an AI LOC incident following the Containment phase of Recovery. After major steps have been taken to disrupt the infrastructure necessary to arrest the AI system(s) operating beyond human control, the system(s) is no longer able to operate in an unrestricted manner and the ceiling of damage from the event will have been arrested. However, the AI system will not necessarily be neutralized nor the threat eradicated—the AI system might be dormant or may be confined in its ability to coordinate or take action in the physical world. It is possible and likely that upon walking back of the containment measures, the threat exists again at nearly an equivalent magnitude as prior to the implementation of the infrastructure disruption steps that had successfully led to Containment. Consequently, it is necessary to completely neutralize the threat prior to rolling back the containment measures that resulted in a successful Containment. This process is Threat Neutralization.

Threat neutralization is critical, because rolling back containment measures without first neutralizing the threat essentially resets the clock to the pre-containment conditions.

Implicitly, Threat Neutralization will vary significantly between containment measures using circuit-breakers and containment measures using the escalatory approach; i.e., for accidental and adversarial AI LOC incidents, respectively. For circuit-breakers, the containment measures are designed to cut off and cut on easily and quickly if necessary. However, some steps will still be necessary to neutralize the threats posed that triggered the circuit-breakers. In some cases it may be as simple as shutting down a coordination mechanism that precipitated cascading, dangerous emergent coordination.[22] In other cases it may be far more challenging, and not straightforward, to neutralize threats even for accidental incidents that trigger circuit-breakers. Moreover, the boundary between the simpler cases of neutralization and the more challenging cases of neutralization is likely to be unclear in practice.

Similar to the circuit-breaker approach, automated steps to ensure confidence in neutralization and that we can reset safely should be a goal as much as is possible. When systems do return to operation after circuit-breakers are tripped, they should likely go into a 'safe mode' for diagnostics until complete verification of threat neutralization is possible. Returning to a safe mode state of operations is prudent for all systems after an AI LOC incident, and AI datacenters, and possibly other compute resources, may benefit from custom bootable safe modes that are specific to AI LOC.

Threat Neutralization for adversarial AI LOC incidents is very much a different animal than Threat Neutralization for accidental AI LOC incidents because not only is the threat present, but the threat is adversarial in nature. Adversarial AI systems that have been contained may still be operational. Even if not operational, it is likely that they will once again be instantiated after containment measures are relaxed, meaning they may actively try to avoid neutralization. Moreover, in being adversarial, they are likely to take steps to make neutralization difficult so as to ensure that they persist/survive a Containment/Threat Neutralization cycle, even if in a very limited form.

As an example of an adversarial agent trying to survive, we might consider that adversarial agents could split and save their weights on a distributed network of consumer edge devices. If this was the case, then the agent could write a program, with sufficient credentials[23] to access the devices where the weights are hidden. The program could collect data to represent the state of the world at any time, and if the data—e.g., news articles, communications—reflected that the world was relaxing containment measures after a serious AI LOC incident, the program would recognize this and generate a prompt. This prompt, designed by the system prior to Containment, would be used to re-instantiate the rogue AI system through the distributed network. The computation may be excessive, with equally excessive latency issues, but the agent would again be operating—with new and valuable knowledge learnt from a failed attempt to break free of human control—if steps were not in place to prevent such a scenario from being realized. Being prepared to prevent adversarial agents from avoiding neutralization in this way will be a particularly challenging problem for security professionals.

[22] The example of OpenClaw and Moltbook from January of 26' offer a good example of emergent coordination (e.g., see 1, 2).

Even if strategic steps are not taken by adversarial agents to actively install dormant or near dormant clones, there are additional challenges that could make Threat Neutralization extremely difficult, particularly when adversarial agents are still operational. Containment measures may or may not require disabling rogue AI agents, and even if disabled, rogue AI agents are likely to be re-instantiated by default as containment measures are relaxed. Thus, this presents a case where you are dealing with a rogue AI agent(s) that is still actively operating beyond human control, albeit in a confined environment that prevents catastrophic harms. Adversarial agents that are aware of active efforts to neutralize them could engage in the behavior that led to their breaking free of human control previously. This could be behaviors such as persuasion (OpenAI 2023), scheming (Meinke et al. 2024), deception (Greenblatt et al. 2024), shutdown resistance (Schlatter et al. 2025), etc.

One potential solution for neutralizing an adversarial rogue AI agent is that of a hunter/killer AI proposed by Vermeer (2025). This would be an AI agent designed specifically to hunt and eradicate rogue AI on the system(s) where it proliferated. However, to be effective, the hunter/killer AI will likely need to be as unconstrained as the rogue AI. Vermeer also proposes a digital vermin AI tool to rapidly colonize digital environments to form an intense resources competition with rogue AI, either helping to identify the rogue AI or hindering its ability to effectively pursue its goals due to the competition for resources.[24] Tools such as these are hypothetical, and additional research would be required to design and implement them, and to explore their viability in practice. Moreover, Vermeer proposes these as measures for Containment, and they should be considered for this role, but given the time sensitive nature of AI LOC incident management, and the need to arrest AI LOC incidents as quickly as possible, we suggest that they play a more practical and valuable role in Threat Neutralization because they are not immediately able to sever or degrade the survival dependencies of AI systems in the same manner as the Containment measures proposed here.

It is worth mentioning that there is one surefire way of ensuring threat neutralization, albeit an extreme, extremely costly approach—to leave some or all of the containment measures in place indefinitely. For example, if network infrastructure, including consumer devices, remained disconnected but power was restored, rogue AI containment may be able to persist.[25] It is likely that in such circumstances the rogue AI would not make itself known, and it could be very tempting to restore the disabled infrastructure. However, replacing the infrastructure—e.g., the network and compute infrastructure—would be able to provide nearly complete confidence in the neutralization of the threat. While this may be the safest path, we find it unlikely that such an approach would be pursued, unless the collateral damage from the AI LOC incident had approached the magnitude of existential.

[23] Obtaining bulk credentials consumer edge devices would require long-term planning skills as well as strong capabilities in cyber offensive skills or persuasion, but this would be precisely the kind of preparations for self-preservation that an adversarial agent could be expected to undertake.

[24] Vermeer (2025) does express uncertainty about digital vermin being able to compete effectively against a rogue AI.

Considering the previous point, Containment alone is sufficient for pursuing Recovery. However, Recovery will be much easier and less painful *if* the rogue AI system(s) is completely neutralized. Anything short of complete neutralization will have significant implications for the likelihood of complete or successful Recovery.

# Recovery

Any AI LOC incident that merits Class 1 or Class 2 Containment measures is going to be extremely costly to society. Recovering from such an incident will pose significant challenges, whether from the direct consequences of the LOC incident itself or from the collateral damage of the Response. As a result, **Recovery is an indispensable element of any complete AI LOC incident management framework, especially for plans that focus on catastrophic AI LOC incidents.**

Recovery will be dependent on the degree of Containment measures enacted to arrest the AI LOC incident. For a Class 0 incident, meriting no significant manipulation of critical infrastructure, Recovery will be more straightforward. However, it would be a mistake to assume that Recovery from Class 0 will be easy or should be disregarded.

While Class 0 Containment measures might not cause catastrophic damage, it is easy to imagine AI LOC incidents that don't merit the collateral damage associated with restrictions on critical infrastructure, but that are damaging either to the economy, to digital infrastructure, and to institutions.[26] As a simple example, we can consider a rogue AI agent operating from a datacenter with access to global networks that is able to create and distribute malware exploiting numerous zero-day vulnerabilities. The agent may succeed at this but the effort is detected, the instance of the agent is terminated, and future evaluations of this model are either ceased or performed in a more secure environment.[27] Consequently, significant damage may result without Class 1 or Class 2 Containment measures being implemented.

Ideally, a damage threshold could be established for each AI LOC incident such that when societal damage exceeds it, Class 1 or Class 2 Containment measures become warranted. The thresholds for Class 1 and Class 2 Containment measures should therefore be on the order of magnitude of the damage that the Class 1 and Class 2 Containment measures themselves might precipitate. Thus, we can assume that the damage from Class 1 Containment measures would be greater than the damage from a Class 0 incident, and that the damage from Class 2 Containment measures would exceed the damage from a Class 1 incident. In other words, the threshold for implementing a given class of Containment should be roughly proportional to the damage that class of Containment itself would cause.

---

[25] Similarly, it may be that power was disconnected to arrest an LOC incident, but networks were not disabled. Prior to restoring power, networks might be disabled, to maintain the state of Containment.

[26] Such damage could make it more difficult to Respond and/or Recovery from future incidents, and to realize a prosperous role for humanity in a future that heavily integrates advanced AI (e.g., Predd et al. 2026).

[27] Similar to RAND's Security Level 4 or Security Level 5 conditions for securing model weight security (Nevo et al. 2024). We reference this framework because there is no framework specific to self-exfiltration of data or securing agent development environments against threat models internal to the system.

For Class 1 and Class 2 incidents, Recovery must address not only the damage from the LOC incident itself but also the additional damage caused by the Containment measures — damage that by definition exceeds what a lower class of Containment would have produced. Implicitly, restrictions on critical infrastructure will cause severe if not catastrophic damage, and such damage would not include the damage incurred by the rogue AI in the LOC incident. Moreover, if Threat Neutralization took a significant amount of time, the damage from the Containment measures would increase, potentially significantly.

Loss of critical infrastructure could lead to catastrophic societal-scale damage of many different forms. Economic damage would be hard to avoid, and would be the most probable form of catastrophic damage. However, supply chains are particularly reliant on digital infrastructure, so any sustained disruptions in digital infrastructure (i.e., days-long disruptions) at either regional, national, or global scales could lead to tremendous supply chain shocks or complete supply chain catastrophe. Historically, we have seen nothing like this sort of catastrophe, because previously global networks—including the supply chains that depend on them—have not exhibited the fragility that they do in the modern economy (Helbing 2013). Thus, supply chain catastrophe is also likely to result from Class 1 or Class 2 Containment measures.

Power generation is the most perilous form of critical infrastructure that has been considered, and could clearly result in societal-scale catastrophic consequences.

Considering the potential for catastrophic consequences of critical infrastructure restrictions on the scale of Class 1 and Class 2 Containment measures described in the preceding paragraph, Recovery is likely to be particularly difficult for AI LOC incidents that rise to the level of Class 1 or Class 2. Therefore, it is essential that plans be in place to ensure Recovery is as straightforward as possible so as to minimize the collateral damage that results from extended periods of Threat Neutralization.

Recovery is not just about restoring systems to their pre-incident state, but about incorporating lessons learned to strengthen the framework going forward. Predd et al. (2026) argue that the global security architecture must be able to manage incidents and learn from each one to strengthen the ecosystem; this applies equally to Recovery from AI LOC incidents, where each incident should inform improved Containment, Resilience, and Recovery strategies. Moreover, their emphasis on preserving strategic optionality suggests that Recovery quality should be assessed not only by how quickly pre-incident conditions are restored, but by whether the Recovery preserves humanity's capacity to adapt its approach—a recovery that restores infrastructure but precludes future options or concentrates power would be inadequate. In other words, Recovery must be cautious to avoid complicating or compromising future Response.

Much of Recovery is tied to Resilience, in that inadequate Resilience and emergency preparedness increase the costs and shift the burden to Recovery. If there is sufficient Resilience, then Recovery would be able to focus on restoration of the greatest importance, i.e., governance, digital infrastructure, financial institutions, and supply chains.

For extreme AI LOC incidents and Containment measures[28] that result in some degree of catastrophe, risking societal collapse, existential risk can be reduced by expediting recovery (Pilditch et al. 2024), i.e., implementing Resilience measures to make Recovery quicker. The quality of the Recovery, including ensuring epistemic continuity and restoration of democratic governance is also influential in reducing x-risk (Belfield 2023). Thus, Resilience can further reduce existential risk even in the event of extreme global catastrophes from AI.

# Resilience

The idea of humans losing control to intelligent machines is an idea as old as artificial intelligence or computer science.[29] The fundamental challenge with AI safety is a problem of control (Russell 2019), and AI safety discourse has concentrated almost entirely on prevention and alignment. Despite many ascribing a very high chance of catastrophe from advanced AI (Roose 2024; Carlsmith 2022; Ord 2020), very little work has explored AI LOC incident management because of assumptions that it is unmanageable (Yudkowsky & Soares 2025). However, recent work (e.g., Geist & Moon (2025); Vermeer & Heitzenrater (2025)) suggests that security threats from a superintelligent AI can be minimized with rigorous and robust security engineering measures—i.e., through Resilience.

Resilience is a foundational, albeit overlooked, path to reducing existential risk from AI LOC incidents. It achieves this by increasing Resilience to extreme AI LOC incidents (i.e., those impossible to recover from) and by increasing Resilience to the harm that can result from restricting or shutting down critical infrastructure—i.e., Class 1 and Class 2 Containment measures, respectively. The former reduces existential risk from AI directly, whereas the latter indirectly reduces AI existential risk through increasing the likelihood that decision makers take bold action when necessary despite the harm that the Containment measures will cause.

Unlike proposals that require slowing AI development (which is politically contentious) or solving alignment (which remains technically uncertain), Resilience measures are highly tractable. To understand the difference between the two forms of Resilience, it might help to consider that Resilience to impossible scenarios is about keeping the *ceiling* of AI LOC incidents below the 'impossible' threshold while Resilience to Containment measures is about keeping the *floor* of harm from our own response measures survivable. We further describe the two forms of Resilience below:

1. **Resilience to impossible scenarios:** these efforts focus strictly on maintaining and strengthening the physical and institutional conditions under which Containment will remain possible, regardless of how capable an AI system becomes.
    - **This is an area where immediate, concrete progress can be made through security engineering and infrastructure policy.**

[28] Class 2 measures aren't necessary to precipitate catastrophe leading to some degrees of societal collapse.

[29] Turing describes this in a 1951 paper published posthumously (Turing 1996), but the idea dates as far back as Butler's (1872) Erewhon.

2. **Resilience to Containment measures:** these efforts focus on strengthening societal resilience to the potential harm that restrictions or emergency shutdown of critical infrastructure—i.e., Class 1 or Class 2 Containment measures—would create.
    - **Such efforts may resemble disaster response plans for supplying food and water, but would likely need to be at a national scale.**
    - The Department of Homeland Security is currently tasked with U.S. societal resilience efforts, but it is unclear to the public the extent to which plans exist are robust to restrictions or loss of critical infrastructure at a national scale.

Resilience to Containment measures can have additional positive impacts beyond its potential for reducing AI existential risk. As mentioned in the previous section, Recovery is tied to Resilience in many ways. The most straightforward way to improve Recovery—and thereby increase the likelihood of responsible governance and strong epistemics post-incident—is to increase Resilience to global catastrophes and Containment measures.

## Resilience to Impossible Scenarios

**Fundamentally, the objective of Resilience to impossible scenarios is to reduce the number of impossible to regain control from scenarios. At a high level this is accomplished through two steps:**

1. **Identify all AI LOC scenarios that result in impossible to regain control from circumstances**
2. **To implement Resilience measures that move such scenarios from the 'impossible' to regain control from to 'extremely costly' to regain control from[30]**

**Practically, it is most effective to start with the lowest hanging fruit and move up.** To achieve this, we can utilize existing security engineering practices to dramatically minimize the attack surface available to a superintelligent adversary (Vermeer & Heitzenrater 2025).

A critical insight underpinning this Resilience approach is that we do not need to outsmart a superintelligent system; we need to design systems and procedures that rely on physical and mathematical constraints it cannot circumvent (Geist & Moon 2025). Such constraints form the basis of security protocols and security engineering (Vermeer & Heitzenrater 2025), and they can be used to reduce the attack surface of superintelligent AI systems. **The goal is to minimize the attack surface of deterministic attacks, thus requiring AI systems to pursue attack vectors that are highly uncertain.[31]**

We need to construct an exhaustive list of assumptions necessary to ensure that AI LOC incidents do not become 'impossible' to regain control from. In this list we assume that an AI LOC event has happened, so we do not seek assumptions that prevent AI LOC events from occurring. Further, for the examples below, we assume that currently secure systems are resilient to potentially superintelligent systems (e.g., Vermeer & Heitzenrater 2025). Example assumptions:

[30] The priority is moving from impossible to extremely costly; however, moving further down the ladder from extremely costly to costly or to less costly are reasonable subgoals if we succeed with the primary.

[31] In general, increasing uncertainty through any means of resilience is useful in that it increases the complexity of planning required to subjugate humanity.

- Autonomous robotic systems require periodic cryptographic authorization from human-controlled infrastructure to remain operational, enforced at the hardware/firmware level and not overridable by software
- All large-scale energy generation and distribution retains human-controlled physical rapid shutdown that cannot be bypassed through cyber means alone; authentication for rebooting after shutdown utilizes dual control multi-factor physical authentication

Creating an exhaustive list of assumptions is an exercise, likely best suited to large expert elicitation techniques, and would likely need to be an ongoing process. As mentioned previously, these examples we provide all have limitations, and experts would need to be involved to help sharpen them. Importantly, they are all likely to require much more technical detail, and potentially subelements or sets of assumptions necessary to guarantee a more general and higher-level assumption. Security engineering and threat modeling to identify security primitives and protocols that can serve as the foundation for the security measures to be implemented would also be necessary. Thus, significant resources would be required to robustly pursue Resilience to impossible scenarios, and it is unclear what organizations might be able to manage such an ambitious endeavor.[32]

### Additional Resilience to Impossible Scenario Measures

Additional Resilience to impossible scenarios measures would need to seek to reduce risk further, including from highly uncertain attack vectors such as the manipulation or persuasion of humans by an AI. This includes various possible interventions, but the most significant class of these relates to what is really sensible regulation on dual-use technologies. Resilience targeting such dual-use technologies could effectively reduce the potential attack surface available to rogue AI systems seeking to subjugate humanity. Interventions along these lines would have the additional benefit of reducing the likelihood of accidental AI catastrophes perpetrated by terrorists or other malicious human actors.

Commonly, extreme AI catastrophe scenarios rely on dual use technologies, e.g., genetically engineered pathogens (Kokotajlo et al. 2025; Vermeer et al. 2023) or atomically precise manufacturing (APM; Tegmark 2017; Bostrom 2014). Regulation monitoring production equipment for bioengineering or APM could go a long way in reducing the likelihood that such technologies could be a component in scenarios rising to the level of an AI decisive strategic advantage wherein it is impossible for humans to regain control. Significant legislative efforts related to biorisks have been advanced over the past year, although it is unclear whether any have the political support to become policy.[33]

## Resilience to Containment Measures

Resilience to Class 1 and Class 2 Containment measures are an independent form of Resilience efforts. Minimizing collateral damage from Containment increases the likelihood that such measures are implemented when necessary, which indirectly reduces AI existential risk.[34] Due to this synergy, **it is**

---

[32] Especially if the effort was deemed time-sensitive.

[33] e.g., the AI Action Plan (White House 2025) & the National Security Commission on Emerging Biotechnology report (NSCEB 2025).

[34] As circuit-breakers do not require authorization and trigger based on predetermined thresholds/measures, this benefit is only relevant to adversarial AI LOC incidents and not accidental incidents. Regardless, there is significant value in risk reduction.

**essential that decision makers are comfortable with Resilience to potentially catastrophic harms from Containment measures so that they are confident in taking the necessary actions when it matters.**

Resilience to Containment measures constitutes a different form of Resilience than the security engineering approaches prioritized for impossible incident Resilience. Here, infrastructure and the institutions responsible for managing it and governing must be maximally Resilient to Class 1 and Class 2 degrees of Containment. To ensure that leaders are comfortable making the decision to implement appropriate Containment measures when the time comes, the value proposition of taking even the most extreme Containment measures must unambiguously outweigh the risk of doing nothing; thus, the costs of taking action must be minimized.

Here, some of the most critical elements to consider are power infrastructure and supply chains for food. Feeding humanity in the event of catastrophe is something that has been considered by researchers (Westcombe et al. 2025; Denkenberger & Pearce 2015), but no states have acted on the insights from this research to increase resilience. For example, Jehn et al. (2025) model the impact of global shocks—e.g., global catastrophic infrastructure loss from cyberattacks—on food supply, demonstrating the fragility of supply chains on one of humanity's most significant survival dependencies.

In the context of Containment Resilience, food is probably the lowest hanging fruit due to the extant research on resilience to food supply shocks (Baum et al. 2015) and feeding humans in the event of catastrophes (Denkenberger & Pearce 2015). It is unknown how robust government plans for Resilience are, and whether they are intended to scale to nation-wide crises, but judging on similar government projects, i.e., Continuity of Governance (COG) plans, we may need to be skeptical (Graff 2017)—COG plans share the same structural challenge of planning for civilizational-scale disruption under political constraints.

Additional concerns arise with respect to Resilience to Containment measures, including shocks to pharmaceutical and medical supply chains, or restrictions on power or energy supply.[35] A more thorough discussion is beyond the present scope, but should be a priority for future work.

## Resilience and Deterrence

Increasing resilience to 'impossible' incidents and to AI LOC Containment measures increases the difficulty of planning required for advanced AI systems seeking to usurp human control and subjugate humanity. Notably, this only applies to adversarial AI LOC incidents, but it can effectively act as a deterrent for systems considering such actions.

For the deterrent to be effective, the threat of the Containment measures has to be credible, but, due to potentially catastrophic collateral damage associated with some Containment measures, it may be unlikely that these measures are implemented without Resilience. Therefore, undertaking appropriate Containment Resilience efforts can increase the likelihood that even the most extreme Containment measures are deployed when they are necessary.

[35] Natural gas supply could be particularly vulnerable in cold nations or states during peak winter months.

As a result, both Resilience to impossible incidents and Containment Resilience—if implemented adeptly—have the second order effect of further reducing catastrophic and existential risk from advanced AI through deterrence.

# Conclusion

In this paper we introduce the first AI LOC incident management framework and taxonomy that maps classes of AI LOC incidents to proportional responses. Further, we introduce two distinct forms of Resilience that are able to reduce catastrophic and existential risk from AI LOC incidents. However, there are numerous limitations, many opportunities for future work, and salient recommendations for action. We describe a few of these below.

## Limitations

One of the most obvious limitations lies in the lack of inclusion of the Detection and Verification steps of AI LOC incident management. This is certainly an area necessary to address in future work. Additionally, the framework itself is limited in the restrictive definition of AI LOC; however, that definition was selected to prioritize avoiding the worst possible outcomes, and as such we feel this definition is justified. Moreover, Detection and Verification may look very different from other frameworks where the definition of AI LOC incidents is not limited explicitly to catastrophic and existential risks due to 'extremely costly' or 'impossible' incidents. Limiting the scope of Detection and Verification may be prudent for early work on those steps due to the challenges presented by not limiting their scope.[36]

Various other limitations to the framework exist. For example, the class boundaries between Containment measure classes are grey, and the practical implications of this are not discussed. Further, while the accidental/adversarial distinction adds value to thinking about AI LOC incident management, the practical implications for recognizing the difference during Detection is very important but not introduced. Moreover, the distinction between circuit-breaker Containment measures and escalatory Containment measures is underdeveloped, and a practically implementable framework will require significant additional work. Further, the additional core components of the framework—Threat Neutralization, Recovery, and Resilience—would also all benefit from further expansion in future work.

Another of the most significant limitations is the missing discussion of the need for international coordination. AI LOC incidents—accidental or adversarial—involving distributed AI instances would almost certainly not limit their distribution to national borders, i.e., levels of system entrenchment are not minimal. As such, major incidents exhibiting substantive system entrenchment could not be managed effectively by any single nation without cooperation or violations of other nations' sovereignty. This point was not discussed, but has significant implications on escalatory Containment measures as well as circuit-breaker Containment measures. Further, many impossible Resilience efforts may have a limited effect if not implemented widely at a global scale.

---

[36] Thinking of Detection may be easiest if distinguishing between Detection in development, i.e., in an AI development environment, and Detection in the wild, i.e., detection of an adversarial or accidental AI incident that began outside of the development environment. The latter is naturally a far more challenging problem.

## Future Work

A comprehensive discussion of directions for future work are too vast to discuss here, and are beyond the scope of this work. As this is a nascent field of study, we recommend that those working at the forefront of the domain (e.g., those contributing on significant work such as Whittlestone & Hobbs 2026, Boudreaux et al. 2025; CARMA 2025; Geist & Moon 2025; Jeanmaire & Boger 2025; Somani et al. 2025; Stix et al. 2025; Vermeer & Heitzenrater 2025) work collectively to create a research agenda. This is an urgent need for others seeking to contribute, as well as for government agencies who need to grasp the breadth of the challenges presented by comprehensive AI LOC incident management. We hope that this work can help accelerate these and other conversations regarding AI LOC incident management.

## Recommendations for Policymakers and Practitioners

The foremost policy recommendation derives from the notion of resilience to both impossible to regain control from AI LOC incidents and AI LOC containment measures. Resilience against the former requires taking actions to increase the robustness of security engineering practices for critical national security systems and critical infrastructure. This should not be a bipartisan issue and should be politically tractable, and there are indications that this is on the mind of U.S. leaders. Yet, there are concerns that actions to this end could be taken to increase resilience but fail to increase resilience to the degree necessary to defend against the attack surface that advanced adversarial AI systems may present. It is therefore unacceptable for any potential legislative or executive action to these ends to be designed for threat models of human actors or adversaries. Policymakers should be cognizant to ensure that any interventions do not discount the importance of resilience against a threat model of an adversarial AI.

An additional policy recommendation is something related to a particularly actionable intervention concerning Detection: the role advanced AI developers have with respect to AI LOC. AI developers will sit in the driver's seat with respect to detecting behaviors that can be used to contain AI LOC incidents prior to their reaching the point of widespread collateral damage; in fact, without significant regulatory intervention, AI developers are the only parties that are likely to be able to prevent true adversarial AI LOC incidents from passing the Class 0 containment threshold.

Due to the unique position that advanced AI developers occupy with respect to containing AI LOC incidents, it is imperative that developers create and implement their own AI LOC incident management frameworks. Leading AI developers already have frameworks for AI security, i.e., Anthropics Responsible Scaling Policy, OpenAI's Preparedness Framework, and Google's Frontier Safety Framework. AI developers should develop protocols involving the reporting of worrying behavior and internal circuit-breakers for containment that are the earliest AI LOC line-of-defense.